# The Role of Quantum Metastability and the Perspective of Quantum Glassiness in Kitaev's Fractional Spin Dynamics: An NMR Study.

Wassilios Papawassiliou[1]*, Nicolas Lazaridis[2,3], Eunice Mumba Mpanga[4], Jonas Koppe[1], Nikolaos Panopoulos[2], Marina Karagianni[2], Lydia Gkoura[5], Romain Berthelot[4], Michael Fardis[2], Andrew J. Pell[1], and Georgios Papavassiliou[2]*

**Affiliations** [1] Centre de RMN à Très Hauts Champs de Lyon (UMR 5280 CNRS / ENS Lyon / Université Claude Bernard Lyon 1), Université de Lyon, 5 rue de la Doua, 69100 Villeurbanne, France

[2] Institute of Nanoscience and Nanotechnology, National Center for Scientific Research "Demokritos", 153 10 Aghia Paraskevi, Attiki, Greece

[3] School of Applied Mathematical and Physical Sciences, National Technical University of Athens, 15780 Zografou, Athens, Greece

[4] ICGM, Univ Montpellier, CNRS, ENSCM, Montpellier 34095, France

[5] BRC, Technology Innovation Institute, PO Box 9639, Masdar City, Abu Dhabi, UAE

*Corresponding authors: wassilios.papawassiliou@ens-lyon.fr, g.papavassiliou@inn.demokritos.gr

**Abstract**

The suppression of magnetic order and the detection of a half-quantized thermal Hall effect in α-RuCl$_3$ under an external magnetic field have sparked significant debate, whether these phenomena point to spin fractionalization, as posited by the Kitaev quantum spin liquid (QSL) model, or if they arise from a more conventional mechanism in an antiferromagnetically ordered spin state. Here, through $^{23}$Na NMR relaxation measurements on the layered cobaltate Na$_2$Co$_2$TeO$_6$ at two distinct magnetic fields (4.7 and 9.4 Tesla), we provide compelling evidence supporting a variant interpretation. While upon cooling, the NMR relaxation times align with the temperature dependence predicted by Kitaev's fractional spin excitations, below 10 K, a dynamically heterogeneous state is detected with Quantum Spin Glass (QSG) characteristics. In this state, the spin fractionalization dynamics unfold over markedly different time scales, and are characterized by quantum metastability, inducing prolonged equilibration as $T \to 0$. This observation challenges the prevailing view of a primarily homogeneous QSL state, as the ground state of tangible Kitaev systems.

**Introduction**

Fault-tolerant quantum computing, predicated on the braiding of non-Abelian anyons, is considered a groundbreaking technological advancement [1, 2]. A key development in this field was the introduction of the 2-D Honeycomb model by Kitaev [3], which relies on the interaction of frustrated S=1/2 spins through nearest neighbor anisotropic Ising exchange couplings. The exact solution of this model reveals a quantum spin liquid (QSL) ground state, characterized by spin excitations that fractionalize into pairs of Z$_2$ gauge fluxes and Majorana fermions [3]. The non-Abelian statistics governing these excitation modes have spurred numerous studies aimed at detecting these elusive quasiparticles.

Interest in this area was further fueled by the potential of certain 4d and 5d transition metal (TM) Mott insulators, such as α-RuCl$_3$ and α-(Li,Na)$_2$IrO$_3$ [4-8], with octahedral coordination and strong spin-orbit coupling (SOC). These materials exhibit a d$^4$ ($t_{2g}^4$) and d$^5$ ($t_{2g}^5$) low spin configuration, respectively, leading to a j$_{eff}$=1/2 Kramers doublet and enabling an effective Kitaev spin Hamiltonian [4]. This concept was later extended to honeycomb cobaltates, particularly those with the Co$^{2+}$ ion in the high spin 3d$^7$ ($t_{2g}^5 e_g^2$) electron state [9, 10]. However, initial enthusiasm was tempered by the realization that standard non-Kitaev magnetic interactions prevalent in most magnetic TM Mott insulators often overshadow the QSL state at low temperatures. This leads to the emergence of an antiferromagnetically (AFM) ordered state, casting doubts on the attainability of a Kitaev QSL state [11].

Despite these challenges, the scientific community's interest was reignited by the potential for suppressing non-Kitaev interactions and re-establishing the QSL phase through the application of an external magnetic field [12-16]. However, diverse experiments and theoretical works have emerged, with some reports suggesting the coexistence of magnons with an excitation continuum in a strong magnetic field, indicative of a more intricate magnetic ground state than a pure Kitaev QSL [17-23]. Recent thermal conductivity experiments, for instance, have proposed explanations based on thermal phonons or topological magnons to account for the thermal conductivity's temperature dependence under a strong magnetic field [24, 25]. Thus, while theoretical predictions concerning the manifestation of spin fractionalization and a Kitaev QSL phase are robust [26-28], the interpretation of experimental results remains contentious, particularly for materials like Na$_2$Co$_2$TeO$_6$ (NCTO), for which strong Heisenberg AFM interactions and off-diagonal Γ and Γ' terms in the magnetic excitation spectrum [29-31] may obscure the experimental signatures of Kitaev interactions.

Solid-state Nuclear Magnetic Resonance (ssNMR) is widely recognized as a powerful technique for investigating low-energy electron spin excitations in condensed matter systems. By measuring the NMR spin-lattice relaxation rate ($1/T_1$) as a function of temperature, crucial insights have been gained into various phenomena, including the electron density of states (DOS) at the Fermi level in metallic systems [32, 33], electron pairing mechanisms in superconducting materials [34-36], heavy fermion systems [37, 38], and Dirac and Weyl quasiparticle excitations in topological materials [39, 40]. Furthermore, ssNMR is shown to play a vital role in characterizing magnon excitations in both ferromagnetic (FM) and antiferromagnetic (AFM) systems [41, 42]. In the domain of Kitaev QSLs, spin fractionalization, and the presence of Kitaev interactions have been probed using $1/T_1$ experiments on α-RuCl$_3$ [43], prompting significant interest in employing this technique to study related materials such as NCTO, which presents stronger antiferromagnetic interactions compared to its 4d counterpart α-RuCl$_3$.

To gain a deeper understanding of low-temperature electron spin excitations, we conducted variable temperature $^{23}$Na NMR measurements on NCTO powder samples across a wide temperature range (3-300 K) at 4.7 Tesla and 9.4 Tesla. Our findings reveal several notable observations: At 9.4 Tesla, the experimental spin-lattice relaxation rate $1/T_1$ vs. temperature curve aligns exceptionally well with the theoretically predicted curve for the pure Kitaev model [43]. This supports the hypothesis that spin fractionalization in NCTO serves as the primary relaxation mechanism at low temperatures. Surprisingly, even at the lower magnetic field of 4.7 Tesla, where both $1/T_1$ and spin-spin relaxation rates ($1/T_2$) exhibit a sharp divergence at ~25 K, characterizing the PM-AFM phase transition, Kitaev fractional excitations persist. Furthermore, by employing advanced Tikhonov regularization algorithms [39, 44] and sophisticated data filtering techniques, we revealed previously unnoticed NMR relaxation features: Below 10 K, both the $T_1$ and $T_2$ relaxation times, split into two components with distinct characteristics. While the mean values of $1/T_1$ continue to align with the temperature dependence predicted by the Kitaev QSL model [43], the $T_2$ measurements reveal the emergence of ultraslow spin fluctuations that follow an Arrhenius activation law. This kind of dynamic heterogeneity, that is characterized by significantly different time scales, is reminiscent of spin glass formers. Evidently, for T<10 K, spin fractionalization in NCTO and related Kitaev systems is intricately linked to the development of a quantum spin glass state, particularly in the vison branch of the fractional spin excitations, as discussed further below. The fundamental aspects of the spin dynamics revealed in this study are summarized in the graphical representation of Fig. 1c.

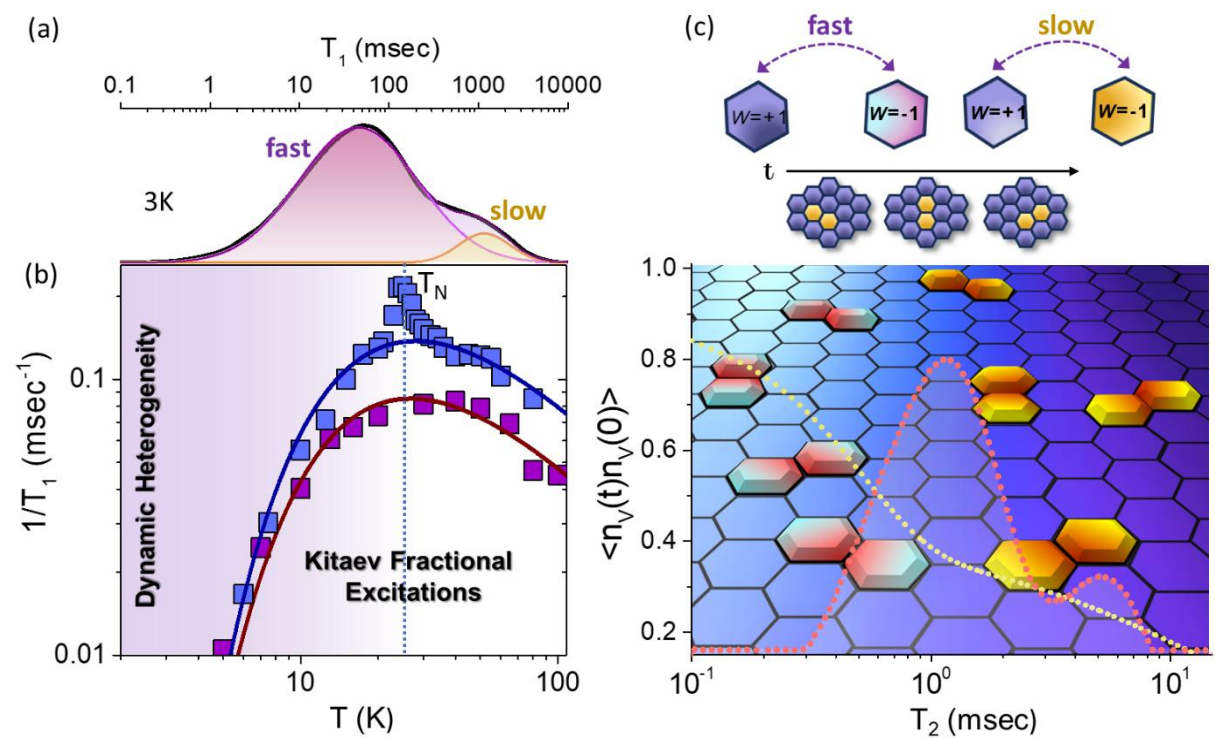

**Figure 1. $^{23}$Na NMR Signature of Kitaev Fractional Spin Excitations.**

**(a)** $^{23}$Na NMR $T_1$ distribution at 3 K in a magnetic field of 9.4 Tesla, spanning over four orders of magnitude. Below 10 K, a second long $T_1$ component emerges, associated with ultra-slow fluctuations following an Arrhenius activation, as elaborated in the article. **(b)** The relaxation rate $1/T_1$ plotted against temperature in magnetic fields of 4.7 and 9.4 Tesla (blue and purple squares, respectively). The divergence of $1/T_1$ at around 25 K in the 4.7 Tesla experiments signifies the presence of a PM-AFM phase transition. Solid lines represent fits to equation 3, highlighting the prevalence of Kitaev spin fractionalization as the primary source of NMR relaxation, even within the nominally "AFM" phase at 4.7 Tesla. **(c)** Graphical representation illustrating key aspects of the spin dynamics discussed in the article. The red dotted line displays the $^{23}$Na NMR $T_2$ distribution at 3 K in a 9.4 Tesla magnetic field, whereas the yellow dotted line represents a hypothesized vison-density autocorrelation function $\langle n_V(t) n_V(0) \rangle$, dictated by the $T_1$ and $T_2$ analysis. In the strong NMR magnetic fields, and at low temperatures, Kitaev fractional excitations are predominantly visons, i.e., flux excitations, described through the plaquette flux operator $W_p$ [3]. In this representation, the ground state is flux-free, with eigenvalue $W_p = +1$ on all plaquettes, whereas by inverting one spin, a pair of $W_p = -1$ flux excitations is generated (brightly colored hexagons). Remarkably, according to the $T_1$ and $T_2$ NMR relaxation measurements, as the temperature drops below 10 K, spin excitations evolve into a highly heterogeneous dynamic state, indicating a vison-density autocorrelation function, decaying across two distinctly different time scales (fast and slow fluctuating visons).

**Results**

**Evidence of spin fractionalization through $^{23}$Na NMR lineshape and $1/T_1$ measurements**

To explore the evolution of the magnetic structure of NCTO under varying temperatures and strong magnetic fields, we conducted $^{23}$Na NMR lineshape measurements across the temperature range of 3 K to 300 K at magnetic fields of 4.7 and 9.4 Tesla. Without an external magnetic field, NCTO adopts an antiferromagnetic (AFM) ground state characterized by the doubling of the unit cell along the a-axis and zigzag AFM ordering along the b-axis, as illustrated in Supplementary Fig. S2. In this state, $^{23}$Na nuclei effectively probe the cobalt magnetic state through the electron-spin dipole hyperfine interaction, $\widehat{H}_{hf} = \hat{I} \boldsymbol{A_{SD}} \hat{S}$, where $\hat{I}$ represents the Na nuclear spin, $\hat{S}$ the electron spin of the Co ions, and $\boldsymbol{A_{SD}}$ is the spin-dipolar hyperfine coupling tensor [45]. Given that $^{23}$Na is a quadrupolar nucleus (nuclear spin I = 3/2), quadrupolar interactions must also be accounted for when interpreting the spectra [46].

Figures 2a and 2b present the temperature-dependent static $^{23}$Na NMR lineshapes under magnetic fields of 4.7 Tesla and 9.4 Tesla, respectively. As temperature decreases, the spectra exhibit asymmetric broadening under both fields, eventually approximating Gaussian lineshapes at the lowest temperatures. Notably, the frequency shift deviates significantly from typical PM/AFM systems. At 4.7

Tesla, the spectra's center of gravity (marked by cyan stars) initially shifts upward, aligning with expectations for a PM system, and subsequently shifts downward as the temperature approaches $T_N \sim$ 25 K. The PM-AFM phase transition is corroborated by the sharp peak in $1/T_1$ at $T_N$ (Fig. 1b), aligning with previous studies [22, 23]. Upon further cooling, an unexpected shift toward higher frequencies is observed. Concurrently, the lineshapes deviate markedly from the typical AFM profile (indicated by the dotted dark purple line in panel 2a), with the Gaussian-like shape suggesting a partial breakdown of AFM spin ordering. At 9.4 Tesla, the situation becomes even more intriguing. The frequency shows minimal temperature dependence initially, followed by a slow decrease as the temperature falls below 100 K. The lack of any divergence in $1/T_1$, as detailed in Fig. 1b, suggests the absence of a PM-AFM phase transition. These findings indicate an unconventional magnetic state quite distinct from ordinary PM or AFM phases as previously documented [17-23]. However, to draw definitive conclusions about the magnetic state and the possible existence of a Quantum Spin Liquid (QSL) state, deeper insights from further NMR relaxation measurements are essential.

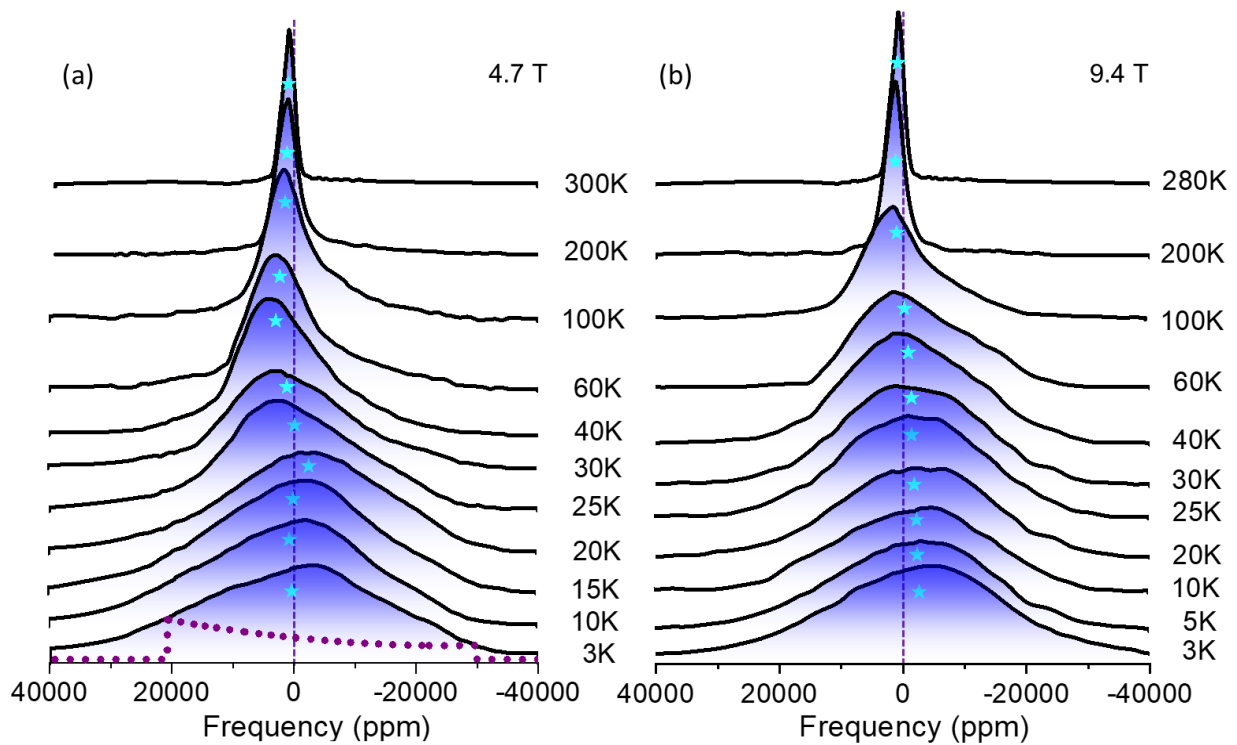

**Figure 2.** $^{23}$**Na NMR Lineshape Measurements - Insights into Spin Disorder at Low Temperatures.** (a) $^{23}$Na NMR lineshape measurements under a 4.7 Tesla magnetic field and (b) under a 9.4 Tesla magnetic field exhibit intriguing temperature dependence, deviating from ordinary PM and AFM phases. The cyan stars indicate the temperature-dependent frequency shift at the spectra's center of gravity. Both fields at low temperatures present experimental NMR lineshapes that markedly differ from anticipated AFM lineshapes (dotted dark purple line in panel 2a) [47].

In the context of 3D AFM insulators like NCTO under weak magnetic fields, the principal spin-lattice $T_1$ relaxation mechanism is expected to be primarily influenced by Raman (two-magnon) process [41, 42, 48]. This process can be described by the formula:

$$\frac{1}{T_1} \propto T^3 \int_{\Delta_m/T}^{\infty} \frac{x\,dx}{e^x - 1} \quad (1),$$

where $\Delta_m$ represents the anisotropy gap in the spin-wave spectrum [48]. In situations when $\frac{\Delta_m}{T} \ll 1$, the integral can be effectively approximated as $\frac{\Delta_m}{T} \exp\left(-\frac{\Delta_m}{T}\right)$, leading to the characterization of gapped magnon excitations in the 3D AFM ordered state by the formula

$$\frac{1}{T_1} \propto T^2 \exp\left(-\frac{\Delta_m}{T}\right) \quad (2).$$

Conversely, in a pure Kitaev system, the theoretical relationship between $\frac{1}{T_1}$ and $T$, calculated using the low-energy limit of the local spin-spin correlation function [33-35], exhibits a distinctive concave shape, accurately represented by the empirical formula [43]:

$$\frac{1}{T_1} \propto \frac{1}{T} e^{-\left(\frac{n\Delta}{T}\right)} \quad (3).$$

Here, $\Delta$ denotes the field-dependent spin excitation gap, and $n$ is a parameter quantifying the degree of fractionalization. Notably, formula (3) shows a maximum at the temperature $n\Delta$, marking the onset of thermally induced gauge flux excitations across the flux gap. Fitting this formula to the theoretical $\frac{1}{T_1}$ vs $T$ curve yields a value of $n = 0.61$ [43]. The distinct temperature dependence of $\frac{1}{T_1}$ between magnon excitations and Kitaev fractional excitations facilitates clear differentiation of the two types of spin excitations.

To investigate the potential emergence of a low-temperature QSL state, $^{23}$Na $T_1$ measurements were conducted in the temperature range 3 K to 300 K. The QSL state is distinguished by a continuum of electron spin excitations, expected to manifest as a continuum of $T_1$ relaxations. To ensure accurate $T_1$ distributions, we employed advanced Tikhonov regularization algorithms in one and two dimensions. Further details on the NMR relaxation measurement procedures and the Tikhonov regularization algorithm are provided in the supplementary information. Fig. 3a illustrates representative $T_1$ distributions within the temperature range of 3 K to 50 K, under a magnetic field of 9.4 T. These distributions exhibit minimal changes down to 20 K, below which they broaden rapidly upon further cooling, revealing the emergence of a second long-$T_1$ component for T < 10 K. The appearance of a wide range of relaxation times below 20 K aligns well with the predicted quantum spin disorder phase below that temperature reported in ref. [15]. To validate the hypothesis of quantum spin disorder, the relaxation rate $1/T_1$ at the center of gravity of the $T_1$ distributions was plotted as a function of temperature in the range of 5 K to 100 K (Fig. 1b). At first glance, $1/T_1$ displays a concave temperature dependence akin to its 4d TM Kitaev counterpart, α-RuCl$_3$, which has been considered as indication of the presence of gapped Kitaev's fractional excitations [43]. This consideration is further supported by the successful fit using equation (3), with a gap value of Δ(9.4 T) ~ 40 K, suggesting NCTO as a potential QSL system. To further elucidate the distinct relaxation mechanisms involved, a comparative analysis of the $\ln(T_1^{-1}T)$ vs $1/T$ curves of NCTO (blue squares) and NNTO ($Na_2Ni_2TeO_6$) (red squares) is presented in Fig. 3b. NNTO is widely accepted to be an ordinary AFM system at low temperatures, governed by magnon excitations for $T < T_N$ [49]. Indeed, despite structural similarities, NNTO aligns well with the two-magnon relaxation mechanism (as indicated by the dashed red line in Fig. 3b, which is fit to equation (2)), while NCTO exhibits a linear decrease (blue solid line) in accordance with equation (3), devoid of any discernible magnon contributions. This clear distinction underscores the prevalence of Kitaev interactions in NCTO.

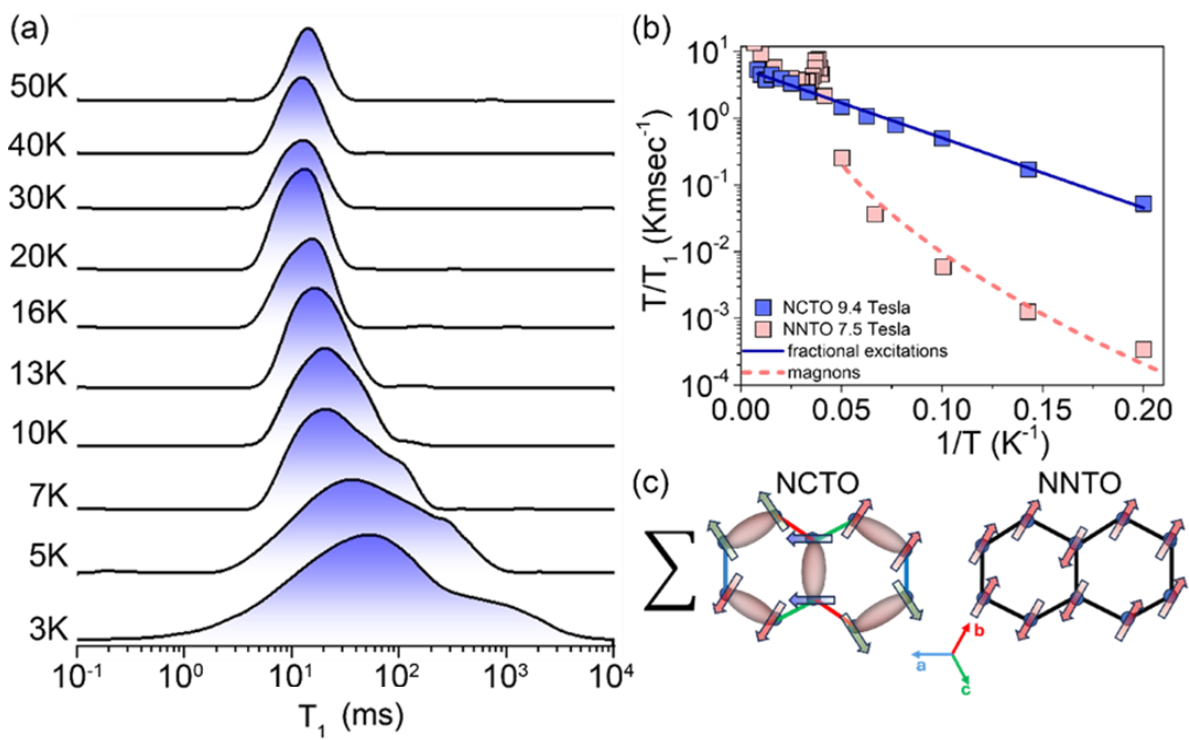

**Figure 3. $T_1$ evidence of a Low-Temperature Quantum Spin Liquid State in NCTO. (a)** $^{23}$Na NMR $T_1$ distributions in the temperature range of 3 K to 50 K in magnetic field of 9.4 Tesla. Below 10 K, the $T_1$ distributions broaden significantly, with a second longer $T_1$ component emerging as the temperature decreases. **(b)** Comparison of the $ln(TT_1^{-1})$ vs $1/T$ curves for NCTO and NNTO: For NCTO, the solid blue line represents an excellent linear fit based on equation (3), providing further evidence of spin fractionalized excitations in the temperature range of 5 K to 25 K. Conversely, for NNTO, the dominance of two-magnon relaxation processes, consistent with equation (2) is observed (dashed red line). The experimental $T_1$ values of NNTO were taken from ref. [49]. **(c)** Schematic representation of the magnetic spin configurations of NCTO and NNTO: In NCTO, the system fluctuates within the degenerate classical ground state manifold (superposition of all Kitaev spin configurations), leading to a highly entangled QSL state with fractionalized excitations. Green, red, and blue bonds in the cartoon

indicate easy Ising bond-axes parallel to the a, b, and c axes, respectively. NNTO showcases the zigzag AFM state.

Notably, despite the divergence at $T_N \sim 25K$, the temperature dependence of $1/T_1$ in the 4.7 Tesla magnetic field (shown by the blue squares in Fig.1b) exhibits a similar concave shape to that observed at 9.4 Tesla, which can be equally well fitted by equation (3). This unexpected behavior, not previously reported in $^{23}$Na NMR studies of NCTO [22, 23], suggests the persistence of Kitaev fractional excitations even inside the nominally AFM phase. It implies coexistence of AFM order with Kitaev fractional spin excitations, at least in the vicinity of $T_N$, consistent with recent muon spectroscopy experiments [50]. By further lowering temperature, Kitaev fractional spin excitations appear to prevail spin dynamics. To validate these results, we compared our $\ln(T_1^{-1}T)$ vs. 1/T analysis with similar plots derived from the experimental data presented in ref. [22]. As evident in Supplementary Figure S5, the $\ln(T_1^{-1}T)$ vs. 1/T plots generated by measurements in magnetic fields of 5 Tesla and 8 Tesla taken from ref. [22], align equally well with the predictions of equation (3), thus enhancing the robustness and validity of our results.

**Observation of Ultraslow Spin Fluctuations at T<10 K through $T_2$ relaxation experiments**

To further elucidate the electron spin dynamics, we have measured the spin-spin relaxation time ($T_2$) over the temperature range of 3 K to 100 K. Given that $^{23}$Na is a quadrupolar nucleus, a quadrupolar Carr-Purcell-Meiboom-Gill (q-CPMG) pulse sequence $((\pi/2)_x - \tau_{cp}/2 - \{(\pi/2)_y - \tau_{cp}\}_N))$ was utilized [51], consisting of a $(\pi/2)_x$ pulse followed by N=300 equidistant $(\pi/2)_y$ pulses, as illustrated in the diagram at the top of Supplementary Fig. S7 as well as at the top of Fig. 5. The measurements presented in Fig. 4 were conducted in a magnetic field of 9.4 Tesla. An illustration of a spin echo decay (SED) signal, captured by recording all read-outs at the midpoint between the $(\pi/2)_y$ pulses, is presented in the inset of Fig. 4a. Subsequently, the $T_2$ distributions were derived by inverting the SEDs using a non-negative Tikhonov regularization algorithm, as depicted in Figs. 4a and 4c. Further information on the experimental setup and inversion methods can be found in Supplementary Figures S3, S4, and the relevant Supplementary Text.

Like the $T_1$ measurements, it is observed that for temperatures T ≤ 10 K, the $T_2$ distributions split into two components (Figs. 4a and 4b). SEDs in this temperature region are characterized by an almost continuous distribution of relaxation times, as depicted in the inset of Fig. 4a, which can be inverted into a bimodal $T_2$ distribution. At higher temperatures, the second long $T_2$ component vanishes, and the $T_2$ distributions are shaped into single-peaked log-Gaussian distributions. An exactly similar behavior is observed in the magnetic field of 4.7 Tesla, as clearly seen in Supplementary Fig. S3. Considering these observations, the $1/T_2$ as a function of temperature in both magnetic fields was plotted, as presented in the inset of Fig. 4c. In the case of the 9.4 Tesla experiments, for T > 10 K, $1/T_2$ (depicted as magenta squares), remains almost unchanged. However, at lower temperatures, $1/T_2$ splits into two branches: one exhibiting an increase in $1/T_2$ with decreasing temperature, while the other demonstrates a rapid decrease in $1/T_2$ upon cooling, resembling the behavior of $1/T_1$. Furthermore, the shorter $T_2$ component experiences significant broadening, as evidenced by the $T_2$ vs. frequency contour plots at temperatures of 3 K and 16 K (Figs. 4b and 4d). The same trend is observed in the 4.7 Tesla experiments, as depicted by the blue squares in the inset of Fig. 4c. In addition to the critical divergence of $1/T_2$ at ~25 K, indicating the transition to the AFM zig-zagged phase, a comparable temperature dependence to the 9.4 Tesla experiments is observed. However, the bifurcation into the two $1/T_2$ components occur at temperatures below 7 K, as distinctly shown in Supplementary Fig. S3. Furthermore, the long $T_2$ component at 3 K exhibits a notably broader profile compared to the 9.4 Tesla results (Supplementary Fig. S3).

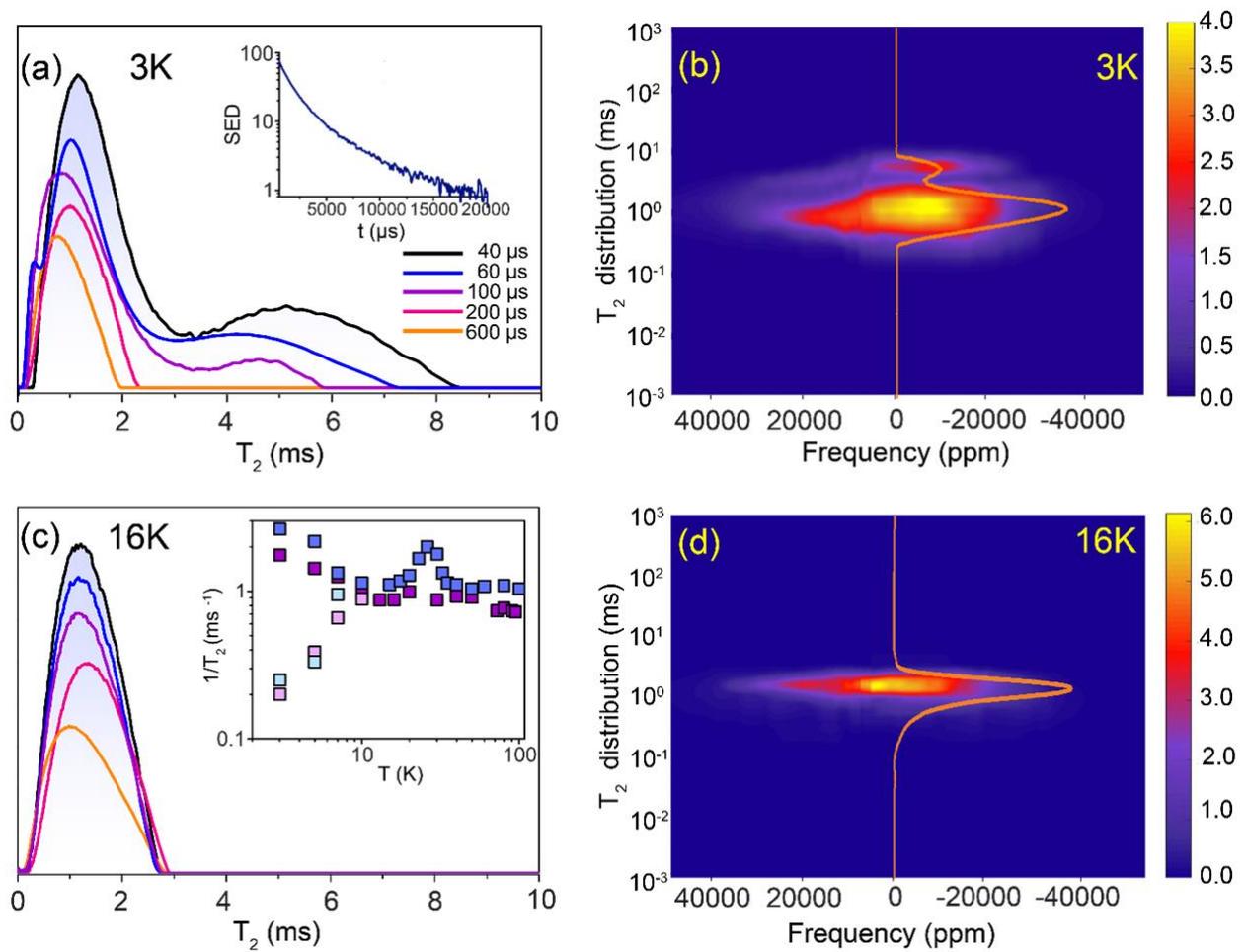

**Figure 4. Insights into ultra-slow spin dynamics through $T_2$ relaxation measurements. (a)** $^{23}$Na NMR $T_2$ distributions in magnetic field of 9.4 Tesla at 3 K for various $\tau_{cp}$ values, obtained by inverting the relevant Spin Echo Decay (SED) curves, derived from a q-CPMG pulse sequence consisting of 300 π-pulses. The inset shows a characteristic SED at $\tau_{cp}$= 40 μs that is inverted to a bimodal $T_2$ distribution. Notably, an increase in $\tau_{cp}$ leads to a rapid decrease and subsequent disappearance of the long $T_2$ component, indicating the presence of extremely slow spin fluctuations as elaborated in the text. **(b)**

The $T_2$ distribution as a function of frequency at 3 K for $\tau_{cp}$= 40 μs. **(c)** The $^{23}$Na NMR $T_2$ distribution at 16 K for various $\tau_{cp}$ values. A unimodal $T_2$ distribution is observed for T > 10 K. No variation in the $T_2$ distributions are noted by changing $\tau_{cp}$. The inset illustrates the $1/T_2$ as a function of temperature at 4.7 Tesla (blue squares) and 9.4 Tesla (magenta squares). At 4.7 Tesla a critical divergence of $1/T_2$ at T ≈ 25 K is observed, indicating the transition to the allegedly low-T AFM phase. Below 10 K the $1/T_2$ bifurcates in two components in both magnetic fields. **(d)** The $T_2$ distribution as a function of frequency at 16 K for $\tau_{cp}$= 40 μs.

One of the most surprising observations that challenges the determination of the q-CPMG experiments is that for T < 10 K both $T_2$ components appear to depend on the time interval $\tau_{cp}$ between the pulses. Particularly, the long $T_2$ component shifts to lower $T_2$ values and diminishes in intensity with increasing $\tau_{cp}$, eventually disappearing for $\tau_{cp}$ > 200 μs, as depicted in Fig. 4a. The dependence of the SEDs on $\tau_{cp}$ is further illustrated in Fig. 5a. However, for T > 10 K it remains independent of $\tau_{cp}$, as evidenced in Figs. 4c and 5b.

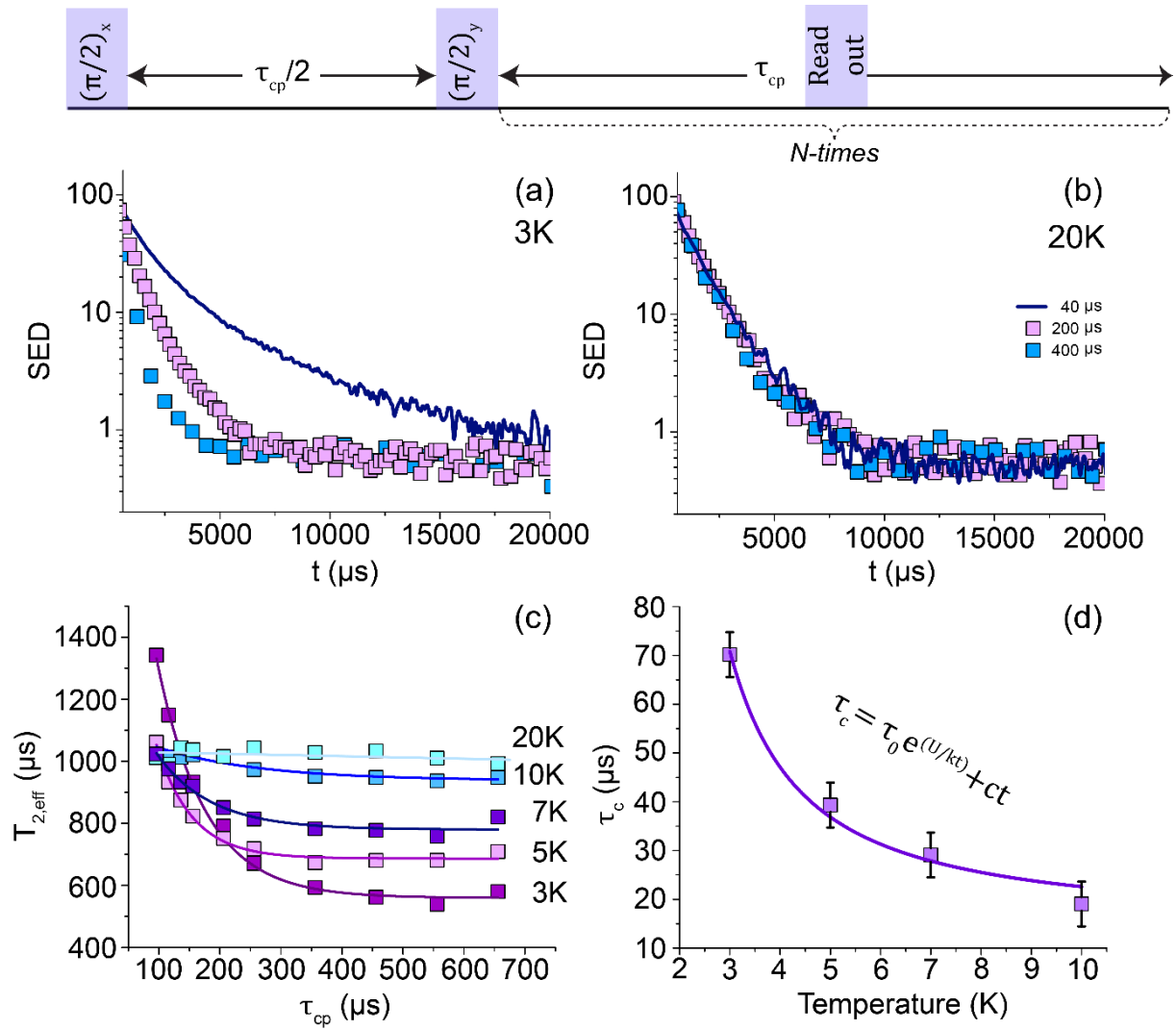

**Figure 5. Arrhenius-like equilibration as an indicator of quantum glassiness. (a)** Spin echo decays (SED) at 3 K in a magnetic field of 9.4 Tesla for different $\tau_{cp}$ values. A schematic illustration of the q-CPMG pulse sequence is depicted at the top of the figure. Increasing $\tau_{cp}$ results in a rapid decrease in the relaxation time $T_2$, indicating very slow spin fluctuations. A similar phenomenon is observed in measurements at 4.7 Tesla, presented in Supplementary Information Fig. S3. **(b)** SEDs at 20 K in a magnetic field of 9.4 Tesla at different $\tau_{cp}$ values. For temperatures above 10 K, SEDs become gradually insensitive to variations in $\tau_{cp}$. **(c)** The effective $T_{2,eff}$, determined by the slope at the first point of the SEDs, as a function of $\tau_{cp}$ at various temperatures. Solid lines represent fits to equation (4), which describes the dependence of $T_2$ on $\tau_{cp}$ in the presence of ultraslow spin fluctuations. **(d)** The correlation time $\tau_c$ obtained from equation (4), as a function of temperature. The solid line is a fit according to an Arrhenius activation law, highlighting the emergence of ultraslow quantum relaxation with glassy characteristics.

To clarify this extraordinary behavior, we have plotted in Fig. 5c the effective spin-spin relaxation time $T_{2,eff}$, determined by the slope at the first point of the SEDs, as a function of $\tau_{cp}$ at various temperatures. Based on the coupling model relaxation theory [52], this single exponential relaxation may be used to describe the dependence of the underlying many-exponentials relaxation mechanism on $\tau_{cp}$. Interestingly, for temperatures below 10 K, the experimental data fit excellently (solid lines) to the expression:

$$T_{2,eff} = [\tau_c \langle S^2(0)\rangle \{1 - (\tau_c/\tau_{cp})tanh(\tau_{cp}/\tau_c)\}]^{-1} \qquad (4),$$

This formula was initially employed to characterize slow chemical exchange of $I = 1/2$ nuclear spins [53] and later adapted to measure the lateral diffusion of quadrupolar nuclei along curved membrane surfaces [54]. In the case of NCTO, the primary source of the $^{23}$Na NMR relaxation are electron-spin fluctuations, mediated by the electron - nuclear hyperfine interactions. By assigning $T_{2,eff}$ to an exponentially decaying autocorrelation function $\langle S(t)S(0)\rangle = \langle S^2(0)\rangle exp(-t/\tau_c)$, which describes the mean electron spin fluctuations, $\tau_c$ in formula (4) represents the correlation time of these fluctuations, and $\langle S^2(0)\rangle$ their square amplitude. Additional details on the implementation of formula (4) to slow fluctuating paramagnetic systems are provided in the Supplementary Information.

The dependence of $T_2$ on $\tau_{cp}$ undeniably indicates the presence of very slow spin fluctuations in the microsecond time scale. To further analyze the origin of this surprising finding, the correlation times $\tau_c$ as a function of temperature for the case of the 9.4 Tesla experiments are displaced in Fig. 5d. A nice fit is achieved with an Arrhenius activation energy law $\tau_c = \tau_0 exp\left(\frac{U}{kT}\right)$, predicting thermally excited spin fluctuations over an energy barrier $\frac{U}{k} = 4.9$ K, and $\tau_0 = 13.82$ μs. Above 10 K no more ultraslow fluctuations are observed. In the case of the 4.7 Tesla experiments, similar behavior is observed, although the slow fluctuations disappear at a sufficiently lower temperature, making it impossible to estimate an activation energy.

To testify how the emergence of $T_2$ branching and its associated ultra-slow spin dynamics correlate with the considerable $T_1$ distribution broadening upon cooling ($T_1$ spreads over 3 orders of magnitude at 3 K), $T_1 - T_2$ correlation spectroscopy experiments were conducted, focusing on the temperature range below 10 K. This approach helps identify the eventual presence of dynamic heterogeneity in the low-frequency spin excitation spectrum.

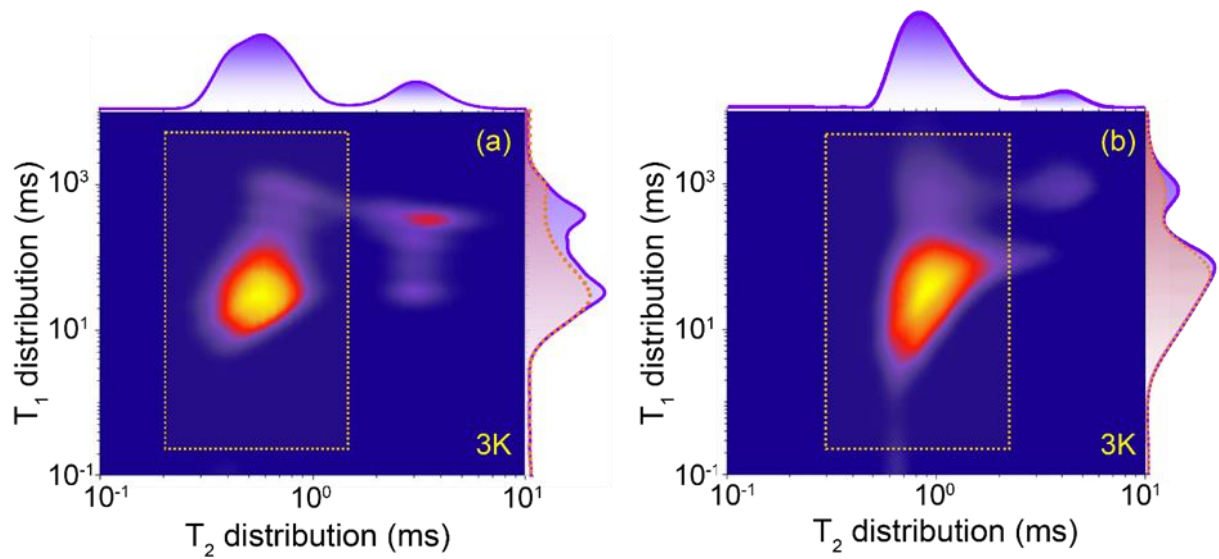

**Figure 6. Dynamic heterogeneity evidenced through two dimensional $T_1 - T_2$ correlation spectroscopy. (a)** The $T_1 - T_2$ contour plot at 3 K under a magnetic field of 4.7 Tesla, and **(b)** at a magnetic field of 9.4 Tesla. In both instances, two diagonally positioned peaks are observed, indicating dynamically segregated regions within the system. The dotted rectangles delineate areas associated with the $T_1$ component primarily influenced by Kitaev interactions. The presence of diffuse rectangularly shaped contours might suggest a form of entanglement between these regions.

Figure 6 presents the 2D $T_1 - T_2$ correlation spectra at a temperature of 3 K, under magnetic field strengths of 4.7 Tesla (left panel) and 9.4 Tesla (right panel). The contour plots highlight two prominent relaxation components positioned diagonally, with blue-colored doubly peaked projections along each axis. At both magnetic field strengths, the short $T_2$ signal component features a continuous $T_1$ distribution, indicated by the red-colored $T_1$ projections of the spectral regions shown in the dotted boxes. The mean $T_1$ values of these distributions vary with temperature, as explained by equation (3). Overall, the contour plots display a diffuse rectangular shape, suggesting the presence of anti-diagonal cross-peaks and a subtle kind of entanglement between the two timely distinct relaxation components. By increasing temperature, the long T$_2$ component gradually diminishes and disappears at temperatures above 10 K (Supplementary Fig. S6). Summarizing the results so far, as the temperature decreases, NMR indicates the presence of fractional spin excitations dominated by Kitaev-like interactions. Below 10 K, the system breaks into two regions, characterized by $T_1$ and $T_2$ values on distinct time scales, and the emergence of extremely slow fluctuations that follow an Arrhenius activation law. This unique behavior can be seen as the quantum equivalent of the freezing process observed in classical glass formers.

**Possible Origin of the detected Quantum Glassiness**

The unexpected separation of fractional spin excitations into two distinct components associated with glass-like freezing, raises questions about the origins of this intriguing phenomenon. Recent studies have highlighted that certain two-dimensional QSL Hamiltonians, such as the Kitaev honeycomb model [3] and the Toric model [1], along with their three-dimensional extensions like the fracton model [55-57], exhibit slow dynamic relaxation rates like those observed in structural glasses [56]. These systems share important characteristics: they lack quenched disorder, involve local spin interactions, can be exactly solved, and feature a degenerate ground state with topological order [57]. Any quantum glassiness present would relate closely to the fractional excitations in these systems.

In the Kitaev honeycomb model, which is considered to describe NCTO [19, 20], fractionalization is manifested by the partitioning of elementary excitations into coupled immobile $Z_2$ gauge flux excitations, known as visons, and mobile gapless Majorana fermions. Flux excitations are expressed through the plaquette flux operator $W_p = \prod_\circ \sigma_i^\mu \sigma_j^\mu$ [3], with $\sigma_{i,j}^\mu$ representing Pauli spin operators at the *i, j* spin site along the *μ* direction. The ground state is flux-free, with $W_p = +1$ on all plaquettes, whereas flipping one spin generates a pair of $W_p = -1$ flux excitations, as shown in Fig. 1c. When an external magnetic field is applied, visons become mobile, dragging mass Majorana fermions with them [58]. Concurrently, the gapless Majorana fermions develop a gap that, according to theory, varies with the magnetic field as ~ $B^3$ [3, 34]. Low temperature excitations in a strong magnetic field are thus primarily defined by flux excitations.

In this framework, vison mobility significantly influences the system's quantum "thermodynamics," since reducing vison density by cooling depends on vison-vison collisions and annihilations. These processes are impeded if visons are immobilized [58]. According to the theory for the 2D Kitaev honeycomb model, as the system cools, vison density $n_V$ decays over time according to $n_V \sim 1/aDt$ [58], where $a$ represents the probability of vison annihilation during collisions and $D$ denotes the vison diffusion coefficient. The equilibration process, although prolonged, is not sufficient for the system to be classified as a quantum glass. Quantum glassiness requires kinetic constraints akin to those in classical glass formers, where equilibration processes follow an Arrhenius activation law:

$\tau_e = \tau_0 \, exp(U/kT)$ [55]. These constraints are crucial for transitioning into a dynamically metastable state of spatially heterogeneous dynamics, as observed in our $T_2$ experiments below 10 K. This behavior is theoretically predicted in 1D and 2D quantum Hamiltonians when excitation hopping decreases below a critical threshold [59].

In this context, a key question is: What constraints on vison motion in a 2D Kitaev Hamiltonian prompt a system, such as NCTO, to reach equilibrium through a quantum glass-like freezing mechanism? While most theoretical predictions focus on a 2D Kitaev model in the isotropic regime ($J_X = J_Y = J_Z$), recent studies have shown that a 2D Kitaev honeycomb model with anisotropic interactions ($J_X = J_Y << J_Z$) under an out-of-honeycomb-plane magnetic field can be modeled by a gapped $Z_2$ toric code [59]. This model yields fractional excitations with severely restricted motion akin to fractons [60]. The emergence of quantum glassiness in this scenario would serve as a signature of spin fractionalization.

From the NMR standpoint, the creation and annihilation of visons directly influence electron spin fluctuations. Thus, within the Kitaev QSL state and the dynamically heterogeneous state below 10 K, the NMR relaxation times $T_1$ and $T_2$ are considered to directly probe vison dynamics. This is reflected through a vison-density autocorrelation function, $\langle n_V(t)n_V(0)\rangle$, which is essentially proportional to the local electron-spin autocorrelation function, $\langle S(t)S(0)\rangle$. An illustration of $\langle n_V(t)n_V(0)\rangle$ in the dynamically heterogeneous state is presented in Fig. 1c, where based on the $T_1$ and $T_2$ measurements, it is expected to decay with correlation times spanning in two different time scales.

In summary, the application of $^{23}$Na NMR on NCTO in magnetic fields of 4.7 Tesla and 9.4 Tesla has revealed that Kitaev fractional excitations are the primary source of NMR relaxation at low temperatures. Notably, there is no indication of NMR relaxation caused by two- or multi-magnon relaxation mechanisms, even in the 4.7 Tesla magnetic field where the supposedly PM-AFM phase transition occurs at $T_N \sim 25$ K. Below 10 K, in both magnetic fields, the distribution profiles of $T_1$ and $T_2$ exhibit significant broadening and split into two distinct components. While the mean $T_1$ values of the two components continue to follow the Kitaev QSL model, the $T_2$ measurements display remarkably slow spin fluctuations, with correlation times following an Arrhenius activation law, akin to the freezing dynamics observed in structural or spin glass systems. One might suggest that at low temperatures, NCTO transitions into a fractional quantum spin glass state, with spin dynamics spanning over different timescales, and exhibiting significant metastability. In this case, the eventual lack of ergodicity prevents the system from reaching thermal equilibrium and accessing the ground state as $T \to 0$. This highlights the importance of considering quantum glassy behavior when studying the ground state and elementary spin excitations in Kitaev systems.

**Sample preparation and characterization.** The $Na_2Co_2TeO_6$ polycrystalline sample was synthesized using a solid-state reaction method. Initially, stoichiometric amounts of $Na_2CO_3$ (Alfa, 99.95%), $Co_3O_4$ (prepared by calcining $CoCO_3$ (Alfa Aesar, 99%) at 400°C for 15 hours), and $TeO_2$ (Thermo Scientific, 99.99%) were mixed with a 5% excess of $Na_2CO_3$ and thoroughly ground. The mixture was then loaded into a gold boat, inside an alumina crucible and sintered at 900 °C in air for 12 hours. The resulting product was then ground into a fine powder and its phase purity was confirmed using X-ray diffraction (XRD) analysis.

**XRD.** XRD spectra were obtained with an analytical PANalytical X'Pert PRO powder diffractometer, utilizing Cu-Kα radiation (λ = 1.5418 Å) with an applied voltage of 40 kV and intensity of 30 mA. The spectra were collected in the angular range (2θ) of 5–80° with a step size of 0.03°. The obtained XRD data was further analyzed through Rietveld refinement using the FULLPROF software program, providing structural parameters and confirming the hexagonal crystal structure of $Na_2Co_2TeO_6$ in the space group P6322. The unit cell parameters were determined to be a = b = 5.274 Å, c = 11.239 Å, and the volume of the unit cell was calculated to be 270.7701 $Å^3$. The XRD analysis confirmed the formation of a pure $Na_2Co_2TeO_6$ phase, with diffraction peaks indexed to indicate the crystalline structure's integrity.

**NMR measurements.** The frequency-sweep $^{23}Na$ NMR spectra were acquired on two home-built NMR spectrometer under static conditions, equipped with two superconducting magnets operating at magnetic fields 4.7 Tesla and 9.4 Tesla, respectively. For the spin-lattice relaxation time $T_1$ and the coherence lifetime $T_2$ experiments a π/2-t-π/2 saturation recovery pulse sequence and a q-CPMG pulse sequence π/2-$τ_{cp}$/2-{π/2-$τ_{cp}$-π/2}$_N$ with a train of N=300 π/2 pulses and appropriate phase cycling were implemented. The $T_1$ and $T_2$ distribution analysis was performed by applying a non-negative Tikhonov Regularization Algorithm. Given that $^{23}Na$ is an I=3/2 quadrupolar nucleus, analysis of the saturation recovery curves by selectively irradiating the central transition (+1/2 ↔ -1/2) was performed based on the relation $(M_0 - M)/M_0 = \{0.1\exp(-t/T_1) + 0.9\exp(-6t/T_1)\}$ [21, 49]. More details on the relaxation analysis are provided in the Supplementary Information.

**Author contributions.** W.P., and G.P. conceived and designed the experiments and did most of the paper writing (with contribution from all coauthors). Frequency sweep NMR and relaxation experiments were performed by N.L., M.F., M.K., N.P, and L.G. NMR relaxation analysis was performed by W.P., J.K., G.P., and A.J.P. Sample preparation and characterization were performed by W.P., E.M.M., R.B.

**Acknowledgments.** N.L. would like to thank Professor A. Kontos from the NTUA-Greece for the academic support during his Master Thesis.

**Competing Interests statement.** Authors declare that they have no competing interests.

**Additional Information.** Supplementary Information is included in the SI document. This includes, Supplementary Notes, Supplementary Figs. S1 to S7, and Supplementary Table S1.

**References.**

1. Kitaev, A. Fault-tolerant quantum computation by anyons. *Ann. Phys.* **303**, 2-30 (2003).
2. Nayak, C., et al. Non-Abelian Anyons and Topological Quantum Computation. *Rev. Mod. Phys.* **80**, 3 (2008).
3. Kitaev, A. Anyons in an exactly solved model and beyond. *Ann. Phys.* **321**, 2-111 (2006).
4. Jackeli, G. & Khaliullin, G. Mott Insulators in the Strong Spin-Orbit Coupling Limit: From Heisenberg to a Quantum Compass and Kitaev Models. *Phys. Rev. Lett.* **102**, 017205 (2009).
5. Do, S. H., Park, S. Y., Yoshitake, J., et al. Majorana fermions in the Kitaev quantum spin system α-RuCl$_3$. *Nat. Phys.* **13**, 1079-1084 (2017).
6. Banerjee, A., et al. Neutron scattering in the proximate quantum spin liquid α-RuCl$_3$. *Science* **356**, 1055–1059 (2017).
7. Plumb, K. W., Clancy, J. P., Sandilands, L. J., et al. α−RuCl$_3$: A spin-orbit assisted Mott insulator on a honeycomb lattice. *Phys. Rev. B* **90**, 041112 (2014).
8. Chun, S. H., et al. Direct evidence for dominant bond-directional interactions in a honeycomb lattice iridate Na$_2$IrO$_3$ *Nat. Phys.* **11**, 462-466 (2015).
9. Liu, H. & Khaliullin, G. Pseudospin exchange interactions in d$^7$ cobalt compounds: Possible realization of the Kitaev model. *Phys. Rev. B* **97**, 014407 (2018).
10. Liu, H., Chaloupka, J. & Khaliullin, G. Kitaev Spin Liquid in 3d Transition Metal Compounds. *Phys. Rev. Lett.* **125**, 047201 (2020).
11. Wang, Y., Osterhoudt, G. B., Tian, Y., et al. The range of non-Kitaev terms and fractional particles in α-RuCl$_3$. *npj Quantum Mater* **5**, 14 (2020).
12. Banerjee, A., Lampen-Kelley, P., Knolle, J., et al. Excitations in the field-induced quantum spin liquid state of α-RuCl$_3$. *npj Quantum Mater* **3**, 8 (2018).
13. Kasahara, Y., Ohnishi, T., Mizukami, Y., et al. Majorana quantization and half-integer thermal quantum Hall effect in a Kitaev spin liquid. *Nature Phys.* **559**, 227-231 (2018).
14. Yokoi, T., et al. Half-integer quantized anomalous thermal Hall effect in the Kitaev material α-RuCl$_3$. *Science* **373**, 568-572 (2021).
15. Lin, G., Jeong, J., Kim, C., et al. Field-induced quantum spin disordered state in spin-1/2 honeycomb magnet Na$_2$Co$_2$TeO$_6$. *Nat. Commun.* **12**, 5559 (2021).
16. Gordon, J. S., Catuneanu, A., Sørensen, E. S., et al. Theory of the field-revealed Kitaev spin liquid. *Nat. Commun.* **10**, 2470 (2019).
17. Chern, L. E., Kaneko, R., Lee, H. Y. & Kim, Y. B. Magnetic field induced competing phases in spin-orbital entangled Kitaev magnets. *Phys. Rev. Res.* **2**, 013014 (2020).

18. Li, S., Yan, H. & Nevidomskyy, A. H. Magnons, Phonons, and Thermal Hall Effect in Candidate Kitaev Magnet α-RuCl$_3$. *2023*. Preprint at https://doi.org/10.48550/arXiv.2301.07401

19. Yao, W. & Li, Y. Ferrimagnetism and anisotropic phase tunability by magnetic fields in Na$_2$Co$_2$TeO$_6$. *Phys. Rev. B* **101**, 085120 (2020).

20. Songvilay, M., Robert, J., Petit, S., et al. Kitaev interactions in the Co honeycomb antiferromagnets Na$_2$Co$_2$TeO$_6$ and Na$_3$Co$_2$SbO$_6$. *Phys. Rev. B* **102**, 224429 (2020).

21. Chen, W., Li, X., Li, Y., et al. Spin-orbit phase behavior of Na$_2$Co$_2$TeO$_6$ at low temperatures. *Phys. Rev. B* **103**, L180404 (2021).

22. Kikuchi, J., Kamoda, T., Mera, N., Takahashi, Y., Okumura, K., & Yasui, Y. Field evolution of magnetic phases and spin dynamics in the honeycomb lattice magnet Na$_2$Co$_2$TeO$_6$: $^{23}$Na NMR study. *Phys. Rev. B* **106**, 224416 (2022).

23. Lee, C. H., Lee, S., Choi, Y. S., Jang, Z. H., Kalaivanan, R., Sankar, R. & Choi, K. Y. Multistage development of anisotropic magnetic correlations in the Co-based honeycomb lattice Na$_2$Co$_2$TeO$_6$. *Phys. Rev. B* **103**, 214447 (2021).

24. Kee, H. Y. Thermal Hall conductivity of α-RuCl$_3$. *Nat. Mat.* **22**, 6-7 (2023).

25. Hong, X., Gillig, M., Yao, W., et al. Phonon thermal transport shaped by strong spin-phonon scattering in a Kitaev material Na$_2$Co$_2$TeO$_6$. *npj Quantum Mater.* **9**, 18 (2024).

26. Motome, Y. & Nasu, J. Hunting Majorana Fermions in Kitaev Magnets. *J. Phys. Soc. Jpn.* **89**, 012002 (2020).

27. Yoshitake, J., Nasu, J. & Motome, Y. Fractional Spin Fluctuations as a Precursor of Quantum Spin Liquids: Majorana Dynamical Mean-Field Study for the Kitaev Model. *Phys. Rev. Lett.* **117**, 157203 (2016).

28. Nasu, J. Majorana quasiparticles emergent in Kitaev spin. *Prog. Theor. Exp. Phys.*, ptad115 (2023).

29. Kim, C., et al. Antiferromagnetic Kitaev interaction in $J_{eff}$ = 1/2 cobalt honeycomb materials Na$_3$Co$_2$SbO$_6$ and Na$_2$Co$_2$TeO$_6$. *J. Phys.: Condens. Matter* **34**, 045802 (2022).

30. Sanders, A. L., Mole, R. A. & Liu, J. Dominant Kitaev interactions in the honeycomb materials Na$_3$Co$_2$SbO$_6$ and Na$_2$Co$_2$TeO$_6$. *Phys. Rev. B* **106**, 014413 (2022).

31. Yao, W., Lida, K., Kamazawa, K. & Li, Y. Excitations in the Ordered and Paramagnetic States of Honeycomb Magnet Na$_2$Co$_2$TeO$_6$. *Phys. Rev. Lett.* **129**, 147202 (2022).

32. Papawassiliou, W., Carvalho, J. P., Panopoulos, N., et al. Crystal and electronic facet analysis of ultrafine Ni$_2$P particles by solid-state NMR nanocrystallography. *Nat. Commun.* **12**, 4334 (2021).

33. Kawasaki, Y., et al. NMR study of magnetic excitation in LiVX$_2$ (X = O, S). *J. Phys.: Conf. Ser.*, **320**, 012028 (2011).

34. Hebel, L. C. & Slichter, C. P. Nuclear Relaxation in Superconducting Aluminum. *Phys. Rev.* **107**, 901 (1957).

35. Slichter, C. P., Yoshimura, K., Kosuge, K. Low frequency spin dynamics in undoped and Sr-doped $La_2CuO_4$. *Phys. Rev. Lett.* **70**, 1002 (1993).

36. Kitagawa, S., et al. Nuclear spin-lattice relaxation rate in the superconducting state on $La(Fe_{1-x}Zn_x)AsO_{0.85}$ studied by $^{75}$As-NMR. *J. Phys.: Conf. Ser.* **391**, 012130 (2012).

37. Hashi, K., et al. NMR Study of a Heavy Fermion Compound YbAs. *J. Phys. Soc. Jpn.* **67**, 4260-4268 (1998).

38. Tokunaga, Y., et al. $^{125}$Te-NMR Study on a Single Crystal of Heavy Fermion Superconductor $UTe_2$. *J. Phys. Soc. Jpn.* **88**, 073701 (2019).

39. Papawassiliou, W., Carvalho, J. P., Kim, H. J., et al. Detection of Weyl fermions and the metal to Weyl-semimetal phase transition in $WTe_2$ via broadband high-resolution NMR. *Phys. Rev. Res.* **4**, 033133 (2022).

40. Papawassiliou, W., Jaworski, A., Pell, A. J., et al. Resolving Dirac electrons with broadband high-resolution NMR. *Nat. Commun.* **11**, 1285 (2020).

41. Saito, Y. Magnon-excitation contribution to the interface magnetization in Co/Cu superlattices. *Phys. Rev. B* **51**, 3930 (1995).

42. Hori, F., Kinjo, K., Kitagawa, S., et al. Gapless fermionic excitation in the antiferromagnetic state of ytterbium zigzag chain. *Commun. Mater.* **4**, 55 (2023).

43. Janša, N., Zorko, A., Gomilšek, M., et al. Observation of two types of fractional excitation in the Kitaev honeycomb magnet. *Nature Phys.* **14**, 786–790 (2018).

44. Day, I. J. On the inversion of diffusion NMR data: Tikhonov regularization and optimal choice of the regularization parameter. *J. Magn. Reson.*, **211**, 178-185 (2011).

45. Abragam, A. *The Principles of Nuclear Magnetism.* (Oxford Univ. Press, 1961).

46. Slichter, C. P. *Principles of Magnetic Resonance.* 3rd ed. (Springer, 1990).

47. Yamada, Y. & Sakata, A. An Analysis Method of Antiferromagnetic Powder Patterns in Spin Echo NMR under External Fields. *J. Phys. Soc. Jpn.* **55**, 1751-1758 (1986).

48. Beeman, D. & Pincus, P. Nuclear spin-lattice relaxation in magnetic insulators. *Phys. Rev.* **166**, 359-361 (1968).

49. Itoh, Y. $^{23}$Na nuclear spin-lattice relaxation studies of $Na_2Ni_2TeO_6$. *J. Phys. Soc. Jpn.* **84**, 064714 (2015).

50. Ma, J., Jiao, J., Li, X., et al. Static magnetic order with strong quantum fluctuations in spin-1/2 honeycomb magnet $Na_2Co_2TeO_6$. PREPRINT (Version 2) available at Research Square [accessed 2023-12-20]. https://doi.org/10.48550/arXiv.2312.06284.

51. Davis, J. H., Jeffery, K. R., Bloom, M., Valic, M. I. & Higgs, T. P. Quadrupolar echo deuteron magnetic resonance spectroscopy in ordered hydrocarbon chains. *Chem. Phys. Lett.* **42**, 390-394 (1976).

52. Ngai, K. L. Relaxation and Diffusion in Complex Systems. Springer, (2011).

53. Luz, Z. & Meiboom, S. Nuclear Magnetic Resonance Study of the Protolysis of Trimethylammonium Ion in Aqueous Solution-Order of the Reaction with Respect to Solvent. *J. Chem. Phys.* **39**, 366-370 (1963).

54. Bloom, M. & Sternin, E. Transverse Nuclear Spin Relaxation in Phospholipid Bilayer Membranes. *Biochemistry* **26**, 2101-2105 (1987).

55. Prem, A., Haah, J. & Nandkishore, R. Glassy quantum dynamics in translation invariant fracton models. *Phys. Rev. B* **95**, 155133 (2017).

56. Schoenholz, S. S., Cubuk, E. D., Sussman, D. M. et al. A structural approach to relaxation in glassy liquids. *Nature Phys.* **12**, 469–471 (2016).

57. Chamon, C. Quantum Glassiness in Strongly Correlated Clean Systems: An Example of Topological Overprotection. *Phys. Rev. Lett.* **94**, 040402 (2005).

58. Joy, A. & Rosch, A. Dynamics of visons and thermal Hall effect in perturbed Kitaev models. *Phys. Rev. X* **12**, 041004 (2022).

59. Olmos, B., Lesanovsky, I. & Garrahan, J. P. Facilitated Spin Models of Dissipative Quantum Glasses. *Phys. Rev. Lett.* **109**, 020403 (2012).

60. Feng, S., et al. Anyon dynamics in field-driven phases of the anisotropic Kitaev model. *Phys. Rev. B* **108**, 035149 (2023).

# SUPPLEMENTARY INFORMATION

**The Role of Quantum Metastability and the Perspective of Quantum Glassiness in Kitaev's Fractional Spin Dynamics: An NMR Study**

Wassilios Papawassiliou[1]*, Nicolas Lazaridis[2,3], Eunice Mumba Mpanga[4], Jonas Koppe[1], Nikolaos Panopoulos[2], Marina Karagianni[2], Lydia Gkoura[5], Romain Berthelot[4], Michael Fardis[2], Andrew J. Pell[1], and Georgios Papavassiliou[2]*

**Affiliations** [1] Centre de RMN à Très Hauts Champs de Lyon (UMR 5280 CNRS / ENS Lyon / Université Claude Bernard Lyon 1), Université de Lyon, 5 rue de la Doua, 69100 Villeurbanne, France

[2] Institute of Nanoscience and Nanotechnology, National Center for Scientific Research "Demokritos", 153 10 Aghia Paraskevi, Attiki, Greece

[3] School of Applied Mathematical and Physical Sciences, National Technical University of Athens, 15780 Zografou, Athens, Greece

[4] ICGM, Univ Montpellier, CNRS, ENSCM, Montpellier 34095, France

[5] BRC, Technology Innovation Institute, PO Box 9639, Masdar City, Abu Dhabi, UAE

*Corresponding authors: wassilios.papawassiliou@ens-lyon.fr, g.papavassiliou@inn.demokritos.gr

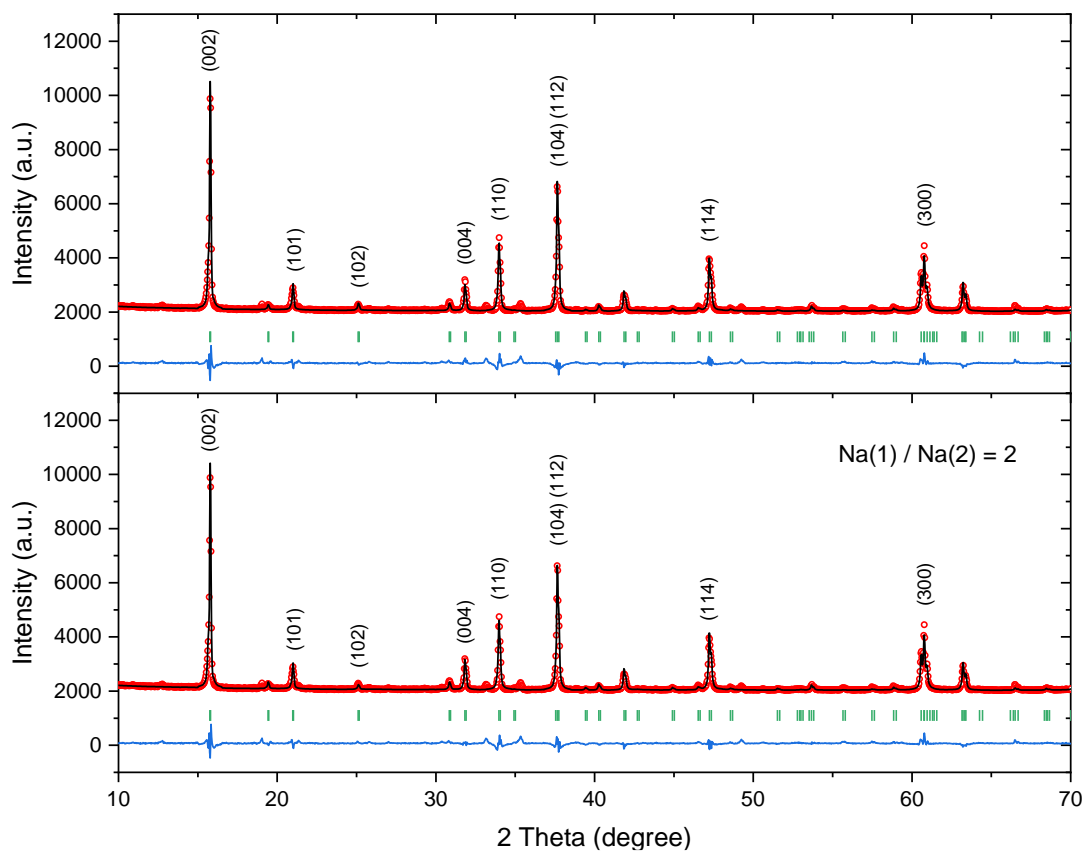

**Supplementary Figure S1**. Observed (circles) and calculated (solid lines) XRD intensities for Na$_2$Co$_2$TeO$_6$ at 296 K with Na occupying equivalent (top) and nonequivalent (bottom) crystallographic positions. Vertical bars show the Bragg peak positions. The difference plot is shown at the bottom.

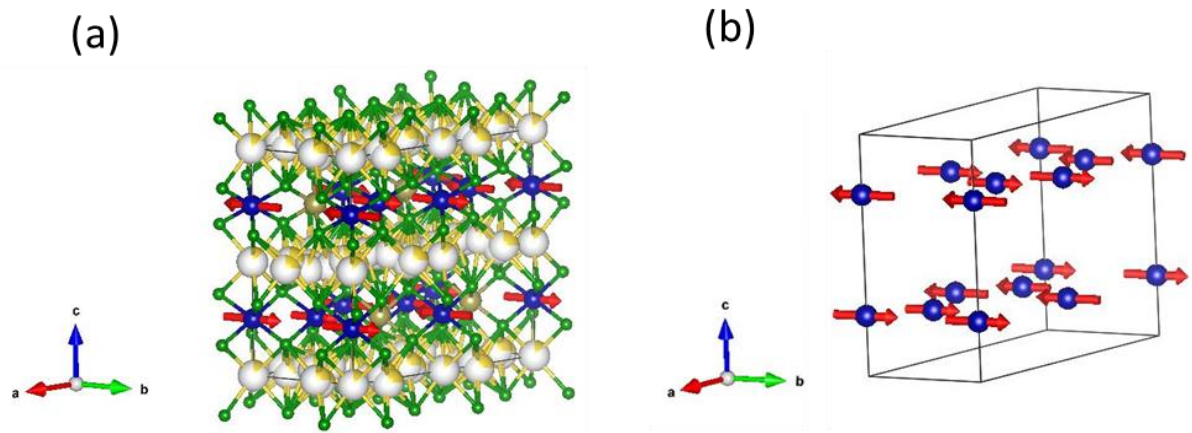

**Supplementary Figure S2**: **The zig-zag AFM state of NCTO**. **(a)** The magnetic structure created according to the mcif file (Bilbao crystallographic server) [1], with the help of the VESTA software. The AFM unit cell is doubled along the a-axis, with spins oriented along the b-axis. Na, Co, Te, and O atoms are presented in yellow, blue, olive-oil, and green color, respectively. Occupation of Na is colored partially white, based on site occupancy. NMR measurements do not confirm this state as the ground magnetic state of NCTO. **(b)** The bare magnetic configuration of the Co atoms in the magnetic unit cell.

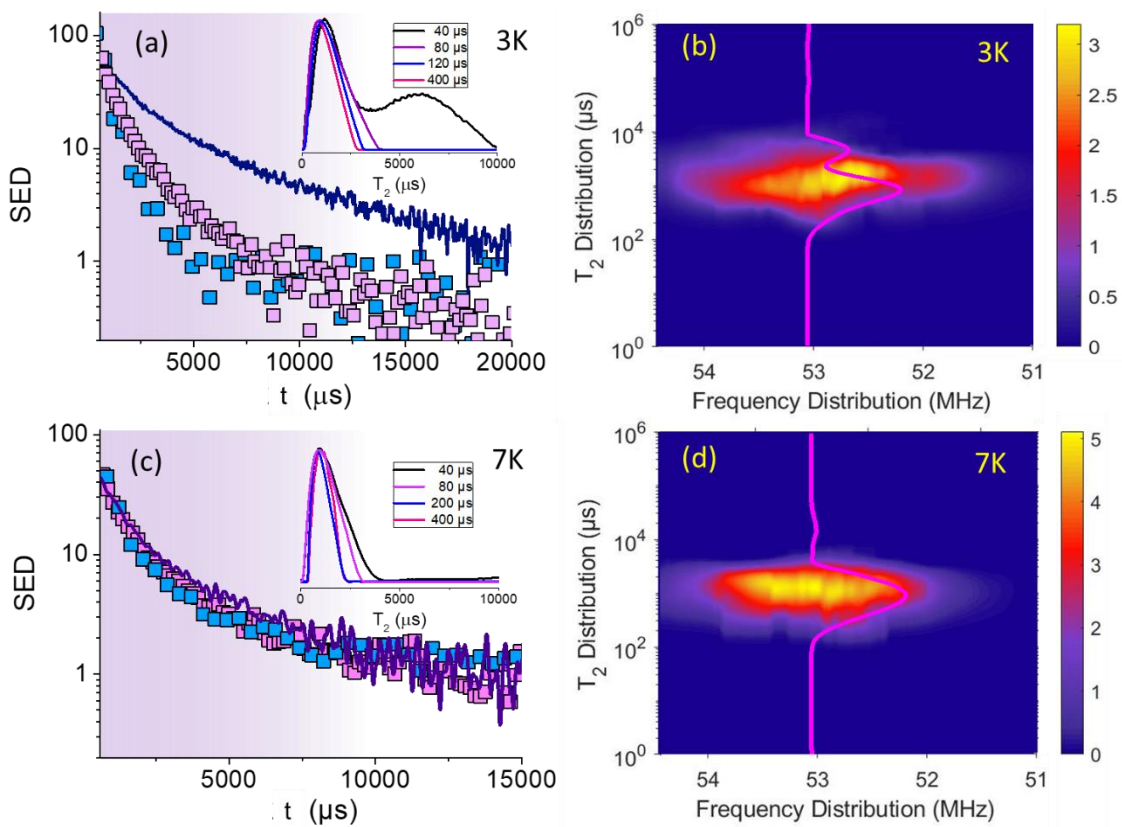

**Supplementary Figure S3: $^{23}$Na NMR $T_2$ dependence on $\tau_{cp}$ and frequency in magnetic field of 4.7T.** SED signal intensity for various $\tau_{cp}$ at (a) 3K and (c) 7K. Inset: $T_2$ distribution obtained by implementing a non-negative Tikhonov regularization algorithm. $T_2$ distribution spectrum at (b) 3K and at (d) 7K. Pink lines are the $T_2$ distributions for $\tau_{cp}$=40μs at frequency ~53.05 MHz.

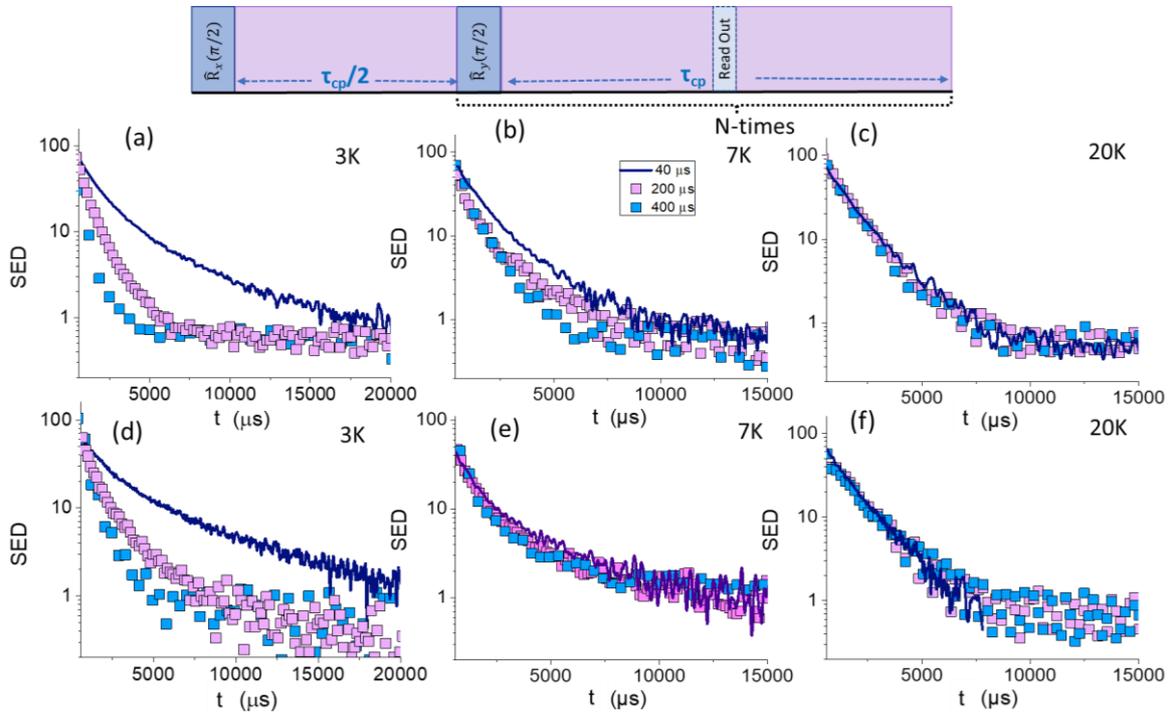

**Supplementary Figure S4: $^{23}$Na NMR Spin Echo Decay (SED) signal intensity obtained by implementing a q-CPMG pulse sequence for various $\tau_{cp}$ values (40, 200 and 400 μs).** Plots (a-c) illustrate the SED in magnetic field of 9.4T at (a) 3K, (b) 7K and (c) 20K and plots (d-f) in magnetic field of 4.7T at (d) 3K, (e) 7K and (f) 20K.

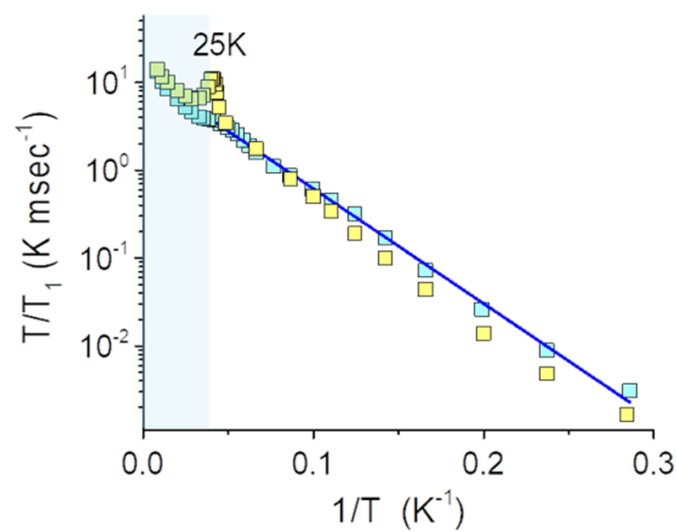

**Supplementary Figure S5: $T/T_1$ vs. $1/T$ analysis on previous experimental data.** $T/T_1$ vs. $1/T$ in magnetic fields of 5 Tesla (yellow squares) and 9 Tesla (cyan squares). Experimental data were taken

from ref. [22] of the main article. The blue solid line is fit according to equation (3) of the main article $\frac{1}{T_1} \propto \frac{1}{T} exp\left(-\frac{n\Delta}{T}\right)$ at 9 Tesla. By considering n=0.61, as explained in ref. [43] of the main article, a gap $\Delta$(9 Tesla) = 44.96 K is predicted.

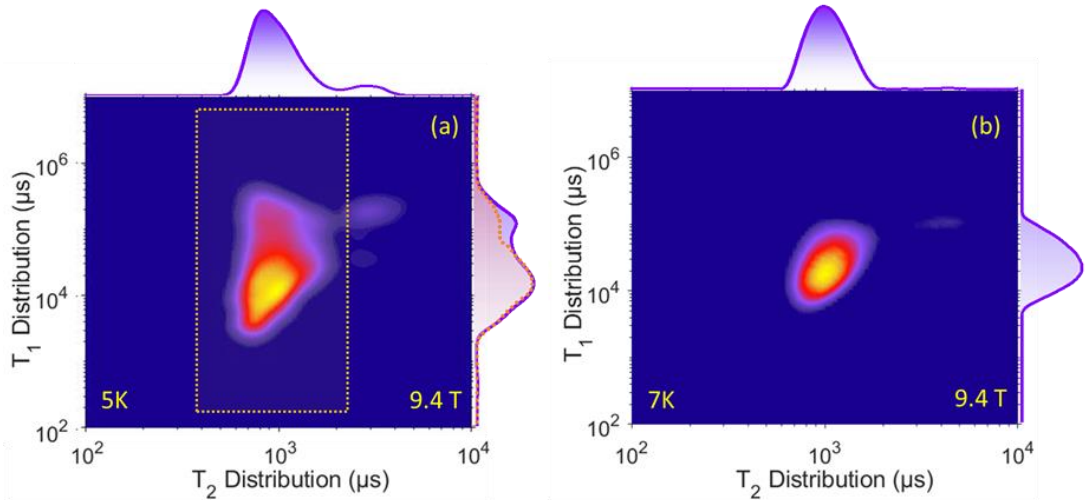

**Supplementary Figure S6: $^{23}$Na NMR $T_1$-$T_2$ correlation spectroscopy spectra in magnetic field 9.4 Tesla.** By increasing temperature, the long $T_2$ component that dominates the spectra at 3K (Fig. 4 in the main article), gradually diminishes and disappears at temperatures above 10 K. Simultaneously, the $T_1$ distribution becomes narrower.

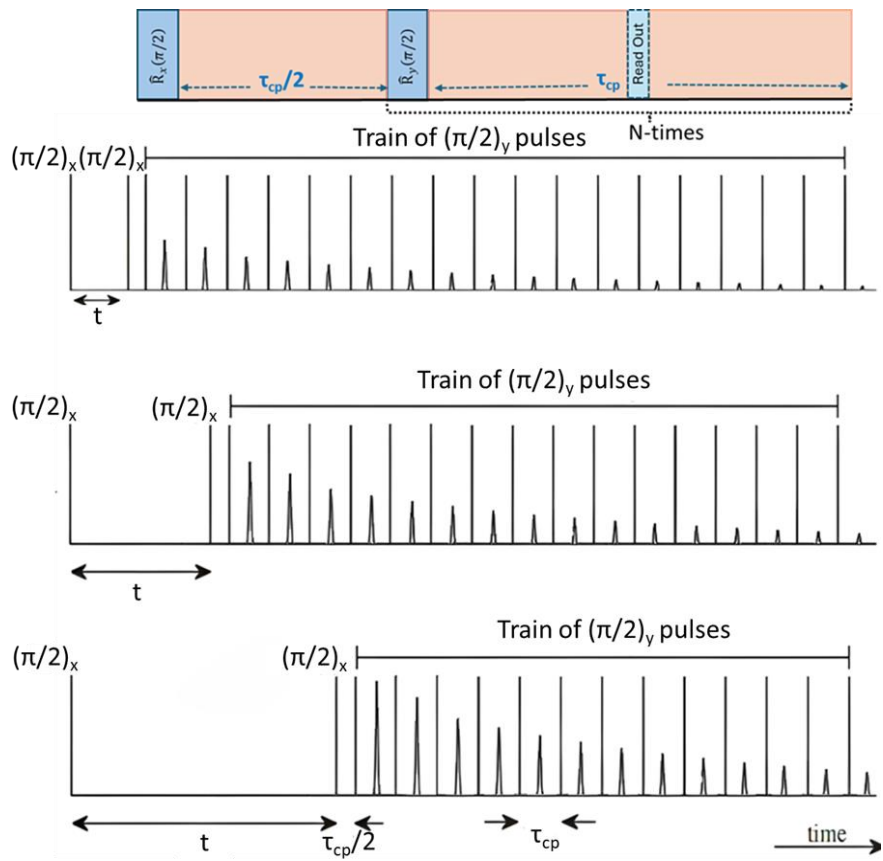

**Supplementary Figure S7: The $T_1$-$T_2$ correlation spectroscopy pulse sequence.** Three distinct instants of the saturation recovery variant of the $T_1$-$T_2$ pulse sequence. The upper panel shows the quadrupolar Carr-Purcell-Meiboom-Gill (q-CPMG) part of the pulse sequence.

| Atom | Wyckoff position | x | y | z | Occ. |
|---|---|---|---|---|---|
| O | 12i | 0.02764 | 0.34203 | 0.66959 | 1 |
| Co(1) | 2b | 0 | 0 | 1/4 | 1 |
| Co(2) | 2d | 2/3 | 1/3 | 1/4 | 1 |
| Te | 2c | 1/3 | 2/3 | 1/4 | 1 |
| Na | 6g | 0.30990 | 0.30990 | 1/2 | 1 |
| $\chi^2 = 0.749$ | | | | | |
| O | 12i | 0.0306 | 0.344 | 0.6727 | 1 |
| Co(1) | 2b | 0 | 0 | 1/4 | 1 |
| Co(2) | 2d | 2/3 | 1/3 | 1/4 | 1 |
| Te | 2c | 1/3 | 2/3 | 1/4 | 1 |
| Na(1) | 6g | 0.38 | 0.38 | 1/2 | 0.67 |
| Na(2) | 2a | 0 | 0 | 0 | 0.33 |
| $\chi^2 = 0.808$ | | | | | |

**Supplementary Table S1:** Refined structural parameters of $Na_2Co_2TeO_6$ in the space group $P6_322$ (No. 182), obtained from powder XRD data at T = 296 K with Na occupying equivalent and nonequivalent crystallographic positions. The unit cell parameters are a = b = 5.2744 Å, c = 11.2388 Å, volume = 270.7701 Å$^3$ (top); a = b = 5.2744 (Å), c = 11.2387 (Å), volume = 270.7695 Å$^3$ (bottom).

**Supplementary Notes**

**Supplementary Note 1:** X-ray diffraction (XRD) was utilized to determine the crystallographic structure of the $Na_2Co_2TeO_6$ compound. The XRD spectra were recorded with an analytical PANalytical X'Pert PRO powder diffractometer. The sample was mounted on a zero-background holder and scanned by using Cu-Kα radiation (λ = 1.5418 Å) with the following experimental conditions: applied voltage of 40 kV, intensity of 30 mA, angular range (2θ) 5–90° and 0.03 steps/s. Rietveld refinement of obtained powder XRD pattern was carried out using the FULLPROF program software [2], [3].

Fig. S1 shows the XRD powder pattern obtained from $Na_2Co_2TeO_6$ at 296 K, along with the refinement indicated with calculated and background difference lines. The purity of the compound was confirmed, and the diffraction peaks were indexed with a hexagonal cell in the space group $P6_322$ (No. 182). The unit cell parameters are a = b = 5.2744 Å, c = 11.2388 Å. The most intense peak is attributed to the (002) plane, which confirms the formation of the hexagonal crystal structure [4-6]. The (002), (101), (102), (004), (110), (112), (104), (114), and (300) lattice planes were found to have 2θ of 15.76°, 20.98°, 25.10°, 31.82°, 33.97°, 37.646°, 37.648°, 47.23°, and 60.78°, respectively, values consistent with those

found in the literature [4-6]. Further structure analysis showed that the Na ions can partially occupy nonequivalent crystallographic positions in different percentages: 67% of the Na reside on sites of symmetry with Wyckoff position index 6g and 33% are in positions with site symmetry 2a (Table S1).

**Supplementary Note 2:** Supplementary Figure S5 presents experimental $1/T_1$ versus temperature data on $Na_2Co_2TeO_6$ powder samples in magnetic fields of 5 Tesla and 9 Tesla, as reported in reference [22] of the main article. The data exhibit a remarkable resemblance to our experimental results, providing additional support for the validity of our findings. Of particular interest are the 5 Tesla measurements, where $Na_2Co_2TeO_6$ is allegedly in an ordinary AFM phase. Like our own experiments, despite the notable divergence at the PM-AFM phase transition, $1/T_1$ follows relationship (3) of the main article, highlighting the prevalence of Kitaev fractional excitations in the NMR relaxation mechanisms.

**Supplementary Note 3:**

In the case of quadrupolar nuclei with $I > 1/2$, the determination of the transverse relaxation time, $T_{2q}$, is typically acquired by employing a two-pulse, $90_x$-$\tau$-$90_y$-$\tau$, quadrupolar echo (qe) pulse sequence [7]. However, this basic sequence fails in capturing slow fluctuations characterized by correlation times that exceed the motional narrowing limit ($\tau_c \gg \tau_M$). In such case, the quadrupolar Carr-Purcell-Meiboom-Gill (q-CPMG) pulse sequence has emerged as a suitable alternative [8]. This pulse sequence, also recognized as the Mansfield-Ware (MW)-4 pulse sequence [9], in its most general form consists of a $90_x$-$\tau_{cp}/2$-$(\varphi_y$-$\tau_{cp}$-$)_N$ pulse train, where $\varphi$ denotes an arbitrary spin-flip angle, as illustrated in the upper panel of Supplementary Figure S7. In a paramagnetic system like NCTO, where interactions between nuclear spins (I) and electron spins (S) primarily drive relaxation, J. S. Blicharski [10, 11] demonstrated that the effective transverse dipolar relaxation rate $1/T_{2q}$ (pseudo-contact relaxation mechanism) caused by fluctuating electron-spin moments is given by:

$$1/T_{2q} = \tau_c \langle S^2(0) \rangle \{1 - (\tau_c/\tau_{cp})tanh(\tau_{cp}/\tau_c)\} \left\{ (1 - \cos\varphi) / \left(1 - \cos\varphi/\cosh\left(\frac{\tau_{cp}}{\tau_c}\right)\right) \right\} \text{ (S1)}$$

The original Blicharski formula has been adjusted to consider the relevant time-dependent part of the hyperfine Hamiltonian, approximated as $H_{hf}(t) = A_{SD}S(t)I_z$. Under this assumption, electron spin fluctuations can be described by an exponentially decaying autocorrelation function $\langle S(t)S(t-\tau) \rangle = \langle S^2(0) \rangle \exp(-\tau/\tau_c)$, which leads to the derivation of Supplementary formula S1, according to the calculation steps in supplementary ref. [12].

By considering $\varphi = 90^o$, formula (4) in the main article is acquired,

$$1/T_{2q} = \tau_c \langle S^2(0) \rangle \{1 - (\tau_c/\tau_{cp})tanh(\tau_{cp}/\tau_c)\} \text{ (S2)}$$

The other two relaxation mechanisms, namely quadrupolar relaxation (via fluctuating electric field gradient) [11], and through-bond hyperfine interactions (Fermi-contact term relaxation), yield the same equation. However, in the case of NCTO, the primary relaxation mechanism is induced by dipolar electron spin fluctuations [13].

**Supplementary Note 4:**

In order to acquire the $^{23}$Na NMR spin-spin relaxation time $T_2$ distribution function $g(T_2)$, the experimental spin-echo decay curves were modelled with a Fredholm integral equation of the first kind [14,15], $\frac{M(t)}{M(0)} = \int_0^{+\infty} k_0(T_2,t) g(T_2) d(log_{10}T_2)$, where $\frac{M(t)}{M(0)}$ is the normalized spin echo decay and $k_0(T_2,t) = exp\left(-\frac{t}{T_2}\right)$. This equation can be transformed in a vector matrix notation to $\boldsymbol{M} = \boldsymbol{K_0 g}$, which in turn can be inverted to obtain the $g(T_2)$ distribution function [14,15]. In the present work, the inversion was achieved by implementing a modified non-negative Tikhonov regularization algorithm [15].

The contour plot in Figures 4b and 4d of the main article, as well as in supplementary Figures S3b and S3d, are made of 50 $g(T_2)$ curves, acquired at 50 consecutive resonance frequencies, covering the whole $^{23}$Na NMR spectra. Ultrasoft pulses ($90^0$~20 μsec) were used so that each time a narrow frequency bandwidth was irradiated.

The spin-lattice relaxation time distribution function $g(T_1)$ was derived using a similar method as $g(T_2)$, by replacing the kernel $k_0(T_2,t)$ in the integral equation with $k_0(T_1,t) = 1 - 0.9 exp\left(-\frac{6t}{T_1}\right) - 0.1 exp\left(-\frac{t}{T_1}\right)$, that is valid for transitions between the $m$ =±1/2 states (central transition), in case of $I = 3/2$ nuclear spins (Refs. [21, 49] of the main article).

Finally, the $T_1$-$T_2$ correlation spectra were obtained by inverting a series of q-CPMG trains, after applying an initial $180^0$ pulse at varying delays from the q-CPMG train, to encode in the $t_1$ direction the $T_1$, as shown in supplementary Figure S7. The series of q-CPMG spin-echo decays were modelled as $\frac{g(t_1,t_2)}{g(0,0)} = \iint_{-\infty}^{+\infty} k_0(T_1,t_1;T_2,t_2) g(T_1,T_2) d(log_{10}T_1) d(log_{10}T_2)$, and subsequently inverted with a 2D Tikhonov regularization algorithm [14, 15]. The kernel $k_0(T_1,t_1;T_2,t_2)$ in the 2D case is the cross product of the two previous kernels for the $T_1$ and $T_2$ inversions.

**Supplementary References**

1.  Lefrançois, E., et al. Magnetic properties of the honeycomb oxide Na$_2$Co$_2$TeO$_6$. *Phys. Rev.* B **94**, 214416 (2016).
2.  Rodríguez-Carvajal, J. Recent advances in magnetic structure determination by neutron powder diffraction. *Physica B: Condensed Matter* **192**, 55–69 (1993).
3.  Rodríguez-Carvajal, J. Recent Developments of the Program FULLPROF. *Newsletter in Commission on Powder Diffraction (IUCr)* **26**, 12–19 (2001).
4.  Xiao, G. et al. Crystal Growth and the Magnetic Properties of Na$_2$Co$_2$TeO$_6$ with Quasi-Two-Dimensional Honeycomb Lattice. *Crystal Growth & Design* **19**, 2658–2662 (2019).
5.  Berthelot, R., Schmidt, W., Sleight, A. W. & Subramanian, M. A. Studies on solid solutions based on layered honeycomb-ordered phases P2-Na$_2$M$_2$TeO$_6$ (M = Co, Ni, Zn). *J. Solid-State Chem.* **196**, 225–231 (2012).

6. Viciu, L. et al. Structure and basic magnetic properties of the honeycomb lattice compounds Na$_2$Co$_2$TeO$_6$ and Na$_3$Co$_2$SbO$_6$. *J. Solid-State Chem.* **180**, 1060–1067 (2007).

7. Davis, J. H., Jeffery, K. R., Bloom, M., Valic, M. I. & Higgs, T. P. Quadrupolar echo deuteron magnetic resonance spectroscopy in ordered hydrocarbon chains. *Chem. Phys. Lett.* **42**, 390–394 (1976).

8. Bloom, M. & Sternin, E. Transverse Nuclear Spin Relaxation in Phospholipid Bilayer Membranes. *Biochemistry* **26**, 2101–2105 (1987).

9. Mehring, M. *Principles of High-Resolution NMR in Solids*. Springer-Verlag, Berlin (1983).

10. Blicharski, S. Nuclear-spin relaxation in the presence of Mansfield-Ware-4 multipulse sequence. *Can. J. Phys.* **64**, 733–735 (1986).

11. Blicharski, S. & Wolak, A. Nuclear Spin Relaxation in periodically perturbed systems. II Like and unlike spins. *Acta Phys. Pol. A* **82**, 511–515 (1992).

12. Lankhorst, D., Schriever, J. & Leyte, J. C. The Time Evolution of the Transverse Magnetization in Carr-Purcell Sequences in the Presence of Chemical Exchange. *J. Magn. Reson.* **51**, 430–437 (1983).

13. Kikuchi, J. et al. Field evolution of magnetic phases and spin dynamics in the honeycomb lattice magnet Na$_2$Co$_2$TeO$_6$: $^{23}$Na NMR study. *Phys. Rev. B* **106**, 224416 (2022).

14. Mitchell, J., Chandrasekera, T. C. & Gladden, L. F. Numerical estimation of relaxation and diffusion distributions in two dimensions. *Prog. Nucl. Magn. Reson. Spectrosc.* **62**, 34–50 (2012).

15. Day, I. J. On the inversion of diffusion NMR data: Tikhonov regularization and optimal choice of the regularization parameter. *J. Magn. Reson.* **211**, 178–185 (2011).